\documentclass[aps,prb,preprint,groupedaddress,showpacs]{revtex4-2}

\usepackage{graphicx}
\usepackage[english]{babel}
\usepackage{amsmath,amssymb,amsfonts,latexsym,amsbsy}
\usepackage{gensymb}
\usepackage{siunitx}
\usepackage[colorlinks=true,linkcolor=blue,citecolor=blue,urlcolor=blue]{hyperref}
\usepackage{xcolor}
\usepackage{lineno}
\usepackage{adjustbox}
\usepackage{upgreek}
\usepackage[normalem]{ulem}
\usepackage{tikz}

\definecolor{lime}{HTML}{A6CE39}
\DeclareRobustCommand{\orcidicon}{
	\begin{tikzpicture}
	\draw[lime, fill=lime] (0,0) 
	circle [radius=0.16] 
	node[white] {{\fontfamily{qag}\selectfont \tiny ID}};
	\draw[white, fill=white] (-0.0625,0.095) 
	circle [radius=0.007];
	\end{tikzpicture}
	\hspace{-2mm}
}

\foreach \x in {A, ..., Z}{\expandafter\xdef\csname orcid\x\endcsname{\noexpand\href{https://orcid.org/\csname orcidauthor\x\endcsname}
			{\noexpand\orcidicon}}
}

\begin{document}


\title{Structural tunability and origin of two-level systems in amorphous silicon}


 \author{H. C. Jacks\orcidA{}}
 \altaffiliation[Present address: ]{Physics Department, California Polytechnic University, San Luis Obispo, California 93407, USA.}
 \affiliation{Physics Department, University of California, Berkeley, California 94720, USA}
 
 \author{M. Molina-Ruiz\orcidB{}}
 \email[Corresponding author: ]{manelmolinaruiz@gmail.com}
 \altaffiliation{H. C. Jacks and M. Molina-Ruiz contributed equally to this work.}
 \affiliation{Physics Department, University of California, Berkeley, California 94720, USA}
 
 \author{M. H. Weber\orcidC{}}
 \affiliation{Physics and Astronomy Department, Washington State University, Pullman, Washington 99164, USA}  

 \author{J. J. Maldonis\orcidD{}}
 \affiliation{Materials Science and Engineering Department, University of Wisconsin -  Madison, Wisconsin 53706, USA}
 
 \author{P. M. Voyles\orcidE{}}
 \affiliation{Materials Science and Engineering Department, University of Wisconsin -  Madison, Wisconsin 53706, USA}
 
 
 \author{T. Metcalf\orcidG{}}
 \affiliation{Naval Research Laboratory, Washington, DC 20375, USA}
 
 \author{X. Liu\orcidH{}}
 \affiliation{Naval Research Laboratory, Washington, DC 20375, USA}
 
 \author{F. Hellman\orcidI{}}
 \affiliation{Physics Department, University of California, Berkeley, California 94720, USA}
 \affiliation{Lawrence Berkeley National Laboratory, Berkeley, California 94720, USA}

\begin{abstract}
Amorphous silicon films prepared by electron beam evaporation have systematically and substantially greater atomic density for higher thickness, higher growth temperature, and slower deposition rate, reaching the density of crystalline Si when films of thickness greater than $\sim300$ nm are grown at 425 \degree C and at $<1$ \AA/sec. A combination of spectroscopic techniques provide insight into atomic disorder, local strains, dangling bonds, and nanovoids. Electron diffraction shows that the short-range order of the amorphous silicon is similar at all growth temperatures, but fluctuation electron microscopy shows that films grown above room-temperature show a form of medium-range order not previously observed in amorphous silicon. Atomic disorder and local strain obtained from Raman spectroscopy reduce with increasing growth temperature and show a non-monotonic dependence on thickness. Dangling bond density decreases with increasing growth temperature and is only mildly dependent on thickness. Positron annihilation Doppler broadening spectroscopy and electron energy loss spectroscopy show that nanovoids, and not density variations within the network, are responsible for reduced atomic density. Specific heat and mechanical loss measurements, which quantify the density of tunneling two-level systems, in combination with the structural data, suggest that two-level systems in amorphous silicon films are associated with nanovoids and their surroundings; which are in essence loosely bonded regions where atoms are less constrained.
\end{abstract}

\date{\today}






\maketitle

\section{Introduction}
\label{S:1}
Amorphous silicon ($a$-Si) is used in photovoltaic, thin-film transistor, quantum computing and gravitational-wave detection technologies~\cite{2000TechnologySilicon, Park2001QuantumNitride, Oliver2013MaterialsBits, Adhikari2020ADetection, Steinlechner2021HowDetectors}, to name a few, since it is much easier to implement than its crystalline counterpart. The performance of amorphous silicon in these technologies is dependent on its properties, which change with the material structure obtained during the preparation process. $a$-Si is typically found 4-fold coordinated, with little variability in bond length, but variability in bond angle and occasional 3-fold coordination leading to dangling bonds. 5-fold coordinated atoms are found in the liquid phase, but not identified in the solid phase~\cite{Fortner1989RadialSilicon, Laaziri1999HighSilicon}. This limited flexibility in solid phase makes it more controllable than systems with highly variable coordination, e.g., amorphous carbon, or multiatomic materials. Four-fold coordination yields an over-constrained atomic environment, which was postulated by Phillips in 1972 to suppress tunneling two-level systems (TLSs)~\cite{Phillips1972}, later confirmed by our work on $a$-Si~\cite{Zink2006, Queen2013, Liu2014, Queen2015JNCS}.

The suppression of TLSs has been of interest to the glass community because they are the source of energy loss at low temperatures and the origin of anomalous acoustic, thermal, and dielectric properties~\cite{Phillips1987}. The ability to control the TLS density in glasses is crucial in applications for which $a$-Si is a potential material, including phase-resonant qubits~\cite{Martinis2005DecoherenceLoss, Gao2007NoiseResonators, Simmonds2009CoherentDefects} and coatings for gravitational-waves detection~\cite{Steinlechner2018Silicon-BasedSensing, Birney2018AmorphousAstronomy}. Electron-beam (e-beam) evaporated $a$-Si shows very low TLS density in high atomic density films, as opposed to high TLS density found in low atomic density films, suggesting that low-density regions are the source of TLSs in e-beam prepared $a$-Si~\cite{Queen2013, Liu2014, Queen2015JNCS}. Therefore, a careful understanding of the structural differences between these films should lead to a deeper knowledge of the nature of the TLSs. It has been shown that coupled dangling bonds create tunneling states on the surface of hydrogen terminated crystalline silicon~\cite{Pitters2011TunnelSurfaces}, but no correlation between dangling bond density and TLS density has been seen in $a$-Si films~\cite{Queen2015JNCS}.

In this work we present extensive structural characterization of $a$-Si films prepared by e-beam evaporation under ultra-high vacuum (UHV) conditions, and show how differences in growth temperature, deposition rate, and film thickness affect the growth mechanisms and film properties. We discuss connections between growth parameters and physical properties, as well as between TLS density and structural features. We conclude that nanovoids are the most likely structures responsible for TLSs in e-beam evaporated $a$-Si films.

\section{Sample preparation and characterization methods}
\label{methods}
Samples were grown by e-beam physical-vapor deposition (PVD) in an UHV chamber with a base pressure $< 5\times10^{-9}$ Torr, and substrate temperature T$_\textrm{S}$ ranging from 50 to 450 \degree C, growth rate r$_\text{t}$ from 0.5 to 2.5 \AA/s, and thickness t from 10 to 750 nm. Samples were grown either onto low-stress amorphous silicon nitride ($a$-SiN$_\textrm{X}$), or onto crystalline silicon with the native oxide layer left intact. A naturally occurring and continuous self-passivating native oxide layer, not thicker than 3 nm~\cite{Morita1990GrowthSurface}, was allowed to form on the surface of most $a$-Si films. In some cases, a thin capping layer of Al was used, which forms an exellent passivation layer on exposure to air that protects the film against oxygen and water subsequent to growth.

Structural characterizations, detailed below, span length scales from $\sim$\,0.1 to 10 nm. These techniques provide information on both the network and structural defects of the films, specifically surface topography via atomic force microscopy (AFM), atomic density via Rutherford backscattering spectrometry (RBS) together with profilometry and AFM, short-range order (SRO) and medium-range order (MRO) via electron microscopy, atomic bond angle deviation and local strain via Raman spectroscopy, dangling bond density via electron paramagnetic resonance (EPR), and empty space or nanovoids volume via positron annihilation Doppler broadening spectroscopy (DBS) measurements. RBS was also used to set limits on impurities concentration. Hydrogen forward scattering (HFS) was used to quantify the films' water concentration.

Substrate curvature measurements were taken using a Tencor FLX-2320 thin-film stress measurement instrument. The stress of $a$-Si films grown on 2-in diameter silicon wafers was determined at room-temperature using Stoney's equation~\cite{Stoney1909} with no corrections, since the thicknesses ratio between films and substrate ($\sim 10^{-4}$) fulfils the condition $\ll 1$~\cite{Klein2000}.

With the aim of looking for depth dependence of atomic density and bond angle deviation, and in particular to investigate the origin of the differences seen in properties of thin and thick films~\cite{Queen2013, Queen2015JNCS}, select thick films were etched in order to study structural properties as a function of depth. Films were wet etched to avoid densification or atomic reconstruction via energy transfer, since a dry etch may alter the samples underlying atomic structure~\cite{Misra1996Etch-inducedReactors}. To improve thickness control and homogeneity, samples were etched under mild sonication in an isopropanol-saturated 6M KOH bath at 80 \degree C for several minutes until specific target thicknesses were reached~\cite{Madou2002}.

\subsection{Ion beam analysis and atomic force microscopy} 
\label{density}
The measured atomic density $n_{at}$ of the $a$-Si films is the combined result of profilometry, RBS and AFM measurements. Profilometry was done using a KLA Tencor ASIQ profilometer, with an error between 0.1\% to 4\% for the thicker and thinner films, respectively. Ion beam analysis was performed by RBS and HFS in an NEC model 5SDH Pelletron tandem accelerator, with $\alpha$ particles beam energy of 3040 keV. Resonant RBS and HFS techniques were used to set limits on impurities, including specifically oxygen and hydrogen. SIMNRA analysis software was used to obtain the samples composition and areal density. A Digital Instruments AFM Nanoscope Dimension 3100 equipped with Budget Sensors Tap300-G silicon probes with a resonant frequency of 300 kHz and force constant of 40 N/m was used to characterize the samples' surface roughness. The scanned areas are $1\times1$ $\upmu$m$^2$ yielding root mean square (RMS) surface roughness $R_q$ and topography of the $a$-Si films, as well as of their substrates. The thickness obtained by profilometry was corrected by the roughness measured by AFM in order to accurately determine the films' volume. Although the roughness was small (less than 2 nm even in the roughest films), this correction is very important for thinner films.

\subsection{Electron microscopy}
\label{microscopyM}
Electron nanodiffraction and electron energy loss spectroscopy (EELS) experiments were performed in a ThermoFisher Titan scanning transmission electron microscope (STEM) operated at 200 kV. The diffracted electron intensity $I$ as a function of the scattering vector $\mathbf{k}$ was measured using energy-filtered electron diffraction on a GIF 865ER with a Gatan US1000 CCD camera at a camera length of 840 mm, an energy filtering with a slit width of 20 eV, and a high coherence, 2 nm diameter nanoprobe beam with a 0.6 mrad semi-convergence angle. One hundred $512\times512$ pixels nanodiffraction patterns were acquired in a $10\times10$ grid of positions $r$ covering a $25\times25$ nm$^2$ area from ten different regions of each sample. The acquisition time for each diffraction pattern was 6 seconds. Each diffraction pattern was averaged over the azimuthal angle to produce the intensity as a function of the scattering vector magnitude $k$.

The nanodiffraction data set was used to evaluate short- and medium-range order. SRO was evaluated from $I(k)$, which is the average of $I(k,r)$ over $r$. This experiment is the equivalent of a large-area diffraction experiment, but with somewhat worse $k$ resolution due to the convergent probe. MRO was evaluated using the fluctuation electron microscopy (FEM) normalized variance $V$, defined as
\begin{equation}
V(k)=\frac{\langle I^{2}(k,r)\rangle}{\langle I(k,r)\rangle^{2}}-1
\end{equation}
where values within $\langle\,\rangle$ are averaged over the position $r$ on the sample. $V$ measures the magnitude of spatial fluctuations in the diffracted intensity, which are sensitive to three and four-body correlation functions~\cite{Voyles2000FluctuationMaterials, Voyles2002FluctuationSTEM, Hwang2012NanoscaleGlass, Zhang2016Long-rangeFilms}. In $a$-Si, MRO spans length scales from approximately the fourth coordination shell ($\sim0.8$ nm) to just under what is detectable by Bragg diffraction ($\sim3$ nm)~\cite{Voyles2003Medium-rangeMicroscopy}. The position in $k$ of peaks in $V$ is controlled by the interatomic spacing inside nanoscale structural heterogeneities (such as deviations in the distribution of ring sizes, see Ref.~\citenum{Zotov1999DependenceDistribution}), and the magnitude of $V$ is controlled by the size, density, and internal order of the heterogeneities. $V$ was corrected for Poisson noise in $I$ as described in ~\cite{Voyles2002FluctuationSTEM}. 

EELS was performed at a camera length of 160 mm, a probe convergence angle of 25 mrad, and an EELS collection angle of 52 mrad. The energy dispersion was 0.05 eV/channel, and the energy resolution was 0.8 eV, measured as the full width at half maximum (FWHM) of the zero-loss peak. EELS experiments focused on the bulk plasmon loss, which is sensitive to the volume number density of electrons in the material, $n$. In the Drude model, the peak plasmon energy $\lambda_0 = \hbar \sqrt{ne^2 / \epsilon_0 m}$, where $e$ is the electron charge, $m$ is the electron mass, and $\epsilon_0$ is the permittivity of free space. At constant composition, shifts in $n$ result from shifts in the volume number density of atoms $\rho$, which has been used, for example, to measure thermal expansion in Al as a means of thermometry~\cite{Mecklenburg2015NanoscaleDevices} and the change in volume of Al on melting~\cite{Palanisamy2011MeltingMicroscope}. Introducing voids into the sample does not shift $\lambda_0$. Instead, it introduces new plasmon modes at different energies associated with the surfaces of the voids. As a result, EELS measurements of $\lambda_0$ report the average density of the sample excluding voids, unlike density measurements by RBS or Archimedes principle, and hence report the average interatomic spacing of the material network.

High resolution transmission electron microscopy (HRTEM) was performed on a FEI Tecnai-TF30 microscope operated at 300 kV. Images with resolution $2048\times2048$ pixels were recorded by a Gatan Ultrascan CCD with 1 second exposure time. $a$-Si films for all TEM experiments were $\sim30$ nm thick grown at different temperatures on $50\times50$ $\upmu$m$^2$ $a$-SiN$_\textrm{X}$ membranes 50 nm thick and measured in plan view.

\subsection{Raman spectroscopy}
\label{BADM}
The distribution of atomic bonds, or bond angle deviation, $\Delta\theta$ and the fractional volume change, or local strain, $tr(\epsilon)$ were determined by Raman spectroscopy performed using an inVia Renishaw micro-Raman/PL system equipped with a 488 nm laser. The laser power was set to $\sim250$ $\mu$W on a spot with area of $\sim2$ $\mu$m$^{2}$, low enough to prevent degradation or crystallization of the amorphous films. Under these conditions, most of the Raman signal comes from the top $\sim30$ nm of the film due to the laser intensity attenuation.

In $a$-Si the bond angles have a variability around the tetrahedral angle of 109.5\degree, the canonical tetrahedral angle in c-Si, characterized by $\Delta\theta$. Beeman et al.~\cite{Beeman1985StructuralSilicon}, and later confirmed by others~\cite{Marinov1997ModelSilicon, Vink2001RamanSi}, established an empirical correlation between $\Delta\theta$ and the FWHM of the $a$-Si transverse optical (TO) peak, observed at $\sim480$ cm$^{-1}$, which yields $\Delta\theta = (\textrm{FWHM} - 15) / 6$. $tr(\epsilon)$ manifests in c-Si Raman spectra through shifts in the TO peak position $\omega$~\cite{Wolf1996Micro-RamanCircuits, Bonera2003CombiningSilicon}. Strubbe et al.~\cite{Strubbe2015StressA-Si:H} demonstrated an empirical relationship between TO peak shifts and local strain in $a$-Si:H, which is valid for isotropic amorphous vibrational mode frequencies. In vibrational density of states calculations, the TO peak vibrations remain almost unaffected by the presence of H atoms, which indicates that atomic interactions are mostly determined by the nature of Si--Si bonds~\cite{Strubbe2015StressA-Si:H}. Experimental Raman spectra of $a$-Si and $a$-Si:H around the TO peak region are quite similar. Therefore, the empirical relationship obtained for $a$-Si:H is also valid for $a$-Si, and establishes $\omega = s~tr(\epsilon) + \omega_0$, where $s = -460\pm10$ cm$^{-1}$ is a fitting parameter and $\omega_0 \approx 480$ cm$^{-1}$ is the bulk $a$-Si TO peak position~\cite{Vink2001RamanSi}.

\subsection{Double-paddle oscillator}
\label{elastic}
Transverse sound velocity ($v_{t}$) measurements of $a$-Si films were taken at 300 mK using the double-paddle oscillator (DPO) technique described elsewhere~\cite{White1995,Liu1998b}. The resonant frequency of the second antisymmetric torsional resonance mode (AS2) at $\sim5500$ Hz is measured to an accuracy of $<$10$^{-5}$ Hz on both the bare and film-laden oscillator. The shear modulus G is measured and $v_{t}$ is then calculated as $v_{t}=\sqrt{G/\rho}$, where $\rho$ is the film mass density.

\subsection{Electron paramagnetic resonance}
\label{DBdensityM}
Electron paramagnetic resonance measurements were used to determine the dangling bond density of the films, using a Bruker ELEXSYS E580 EPR spectrometer with an X-band ER 4123D CW-Resonator at 9.36 GHz. EPR determines the density of dangling bond defects, or unpaired electrons, by measuring the signal strength of the resonant transition between the Zeeman split energy levels of the paramagnetic dangling bond defect. Microwave power (1.5 mW) and magnetic field modulation amplitude (5 G) were adjusted for optimum intensity without line shape distortion. Spectra were measured from 3282 to 3383 G. A bare substrate was used to determine the background contribution, whereas the samples’ spin density $N_S$ was determined by double integration of the experimental absorption first derivative spectra and by comparison to a KCl weak pitch with $N_S$ = 9.5$\times$10$^{12} \pm 5\%$ spins/cm$^3$ and g = $2.0028 \pm 0.0002$. These experimental conditions yielded $N_S$ with a systematic error of $\sim10\%$. The EPR spectra obtained for the $a$-Si samples studied in this work are isotropic with a Land\'e g-factor of 2.0055, typical of $a$-Si dangling bonds~\cite{Thomas1978ElectronStudies}.

\subsection{Doppler broadening spectroscopy}
\label{nanovoidsM}
Positron annihilation experiments using DBS were performed to obtain the volume of empty space, or total volume of nanovoids, as a function of sample depth. Amorphous silicon films were grown on c-Si substrates with the native oxide left intact on the substrate. In order to avoid attenuation from the native oxide on the $a$-Si surface, the films were etched in a buffered oxide etch 10:1 solution for 10 minutes, then introduced into the measurement chamber and brought below $10^{-5}$ Torr within 25 minutes. The vacuum pressure during typical measurements was less than $10^{-7}$ Torr. The positron incident beam energy was varied from 50 eV to 70 keV, which in c-Si yields implantation depths around 1 nm to 15 $\upmu$m, respectively. In the present study, 25 keV yielded an implantation depth in the films' substrate; and data analysis is limited to below 40 keV to avoid systematic error from backscattered positrons that annihilate from the steel vacuum chamber walls.

A summary of DBS measurement and analysis procedures can be found elsewhere~\cite{Tuomisto2013}. Photons emitted by positron annihilation were detected with a high purity germanium detector from EG\&G Ortec, with an energy resolution of 1.45 keV FWHM at the photon energy equivalent to the rest mass of positrons and electrons at 511 keV. To analyze Doppler broadening due to annihilations from nanovoids, we examined the photoelectric peak from 5$\times$10$^{4}$ to 6$\times$10$^{4}$ detected annihilations after suitable background subtraction. The accumulated events $N_{c}$ in a narrow 1.45 keV window around the center and in the two wing regions $N_{w}$ are compared to the total event number in the full photoelectric peak $N_{tp}$. The DBS parameters $S$ and $W$ are the ratios of $N_{c}/N_{tp}$ and $N_{w}/N_{tp}$, respectively. The beam energy $E$ in keV is converted to implantation depth $d$ using the empirical formula $d [nm] = 40~E^{1.6}/\rho$, where $\rho$ is the mass density in g/cm$^{3}~$~\cite{Schultz1988InteractionInterfaces}. VepFit software is used to simulate the empty volume depth profiles optimizing $S$, layer thickness and positron diffusion lengths~\cite{VanVeen1995}. The $S$ parameter yields information about the number of nanovoids and their size, from atomic vacancies to nanovoids
~\cite{Coleman2011}. The $W$ parameter is sensitive to higher momentum electrons and probes the elemental type of the nearest neighbor~\cite{Clement1996AnalysisWing-parameters}.

\section{Results}

In this section we present the results for different properties as a function of growth temperature, growth rate and thickness. We follow the same structure presented in~\ref{methods}, showing first the results that provide insight into the structure and network, and later the structural defects: dangling bonds and nanovoids.

\subsection{Topography and roughness}\label{AFMR}
\begin{figure}
\includegraphics[scale=0.55]{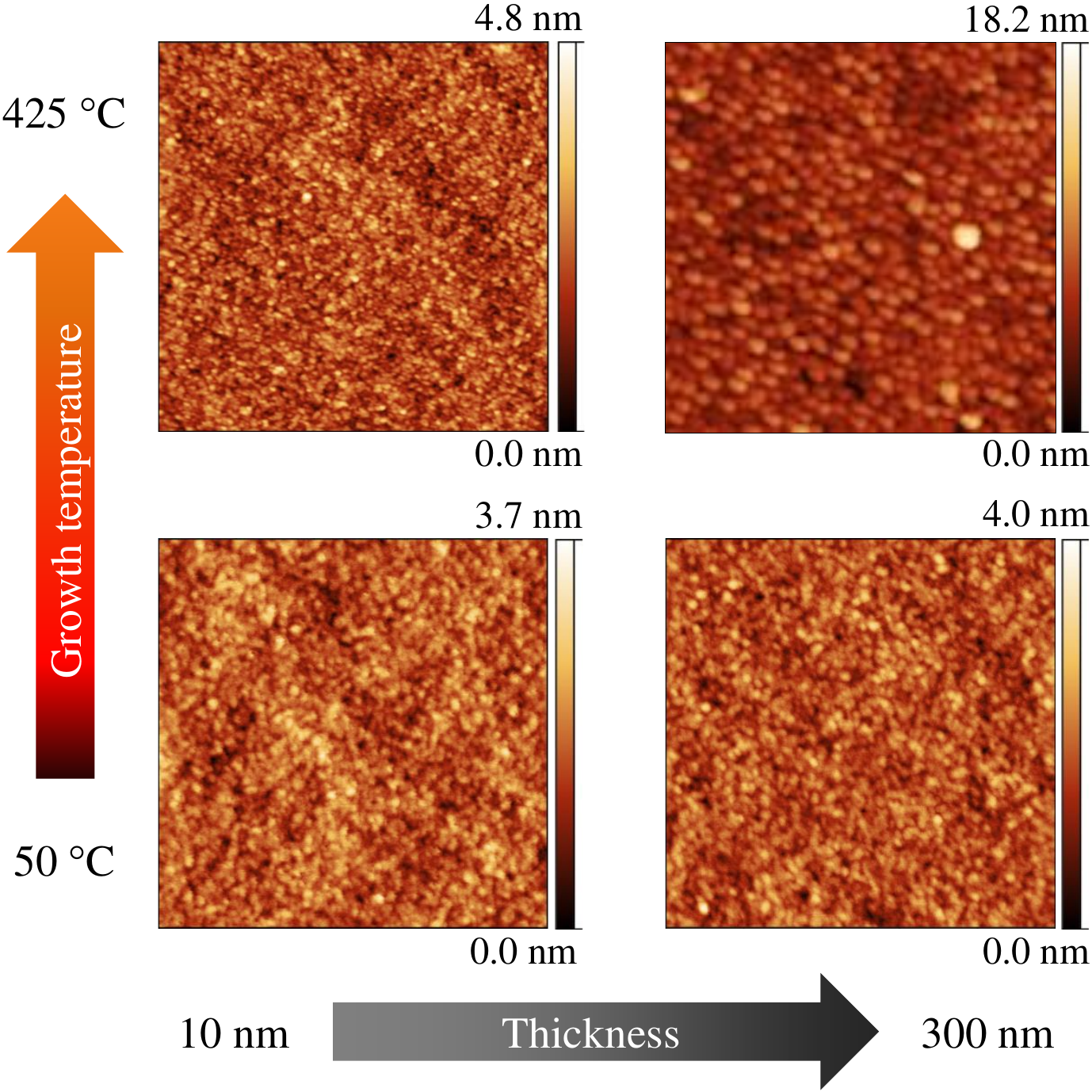}  
\caption{AFM images of the surfaces of a subset of $a$-Si films: 10 and 300 nm thick, left and right, respectively, grown at 50 \degree C and 425 \degree C, bottom and top, respectively. All images are $1\times1$ $\upmu$m$^2$.}
\label{Figure2a}  
\end{figure}

Figure~\ref{Figure2a} shows AFM images of samples grown at different temperatures and thicknesses. These images show the topography of the films' surface, where the roughness is considerably increased for thick films grown at high temperature. The statistical analysis of the images yields an average in plane grain diameter $\sim12$ nm for all films, except for the thick films grown at 425 \degree C that show grains of $\sim20$ nm. These grains measured by AFM are likely to be the end of amorphous columns observed previously in similar $a$-Si films~\cite{Queen2015JNCS}.

\begin{figure}  
\includegraphics[scale=0.95]{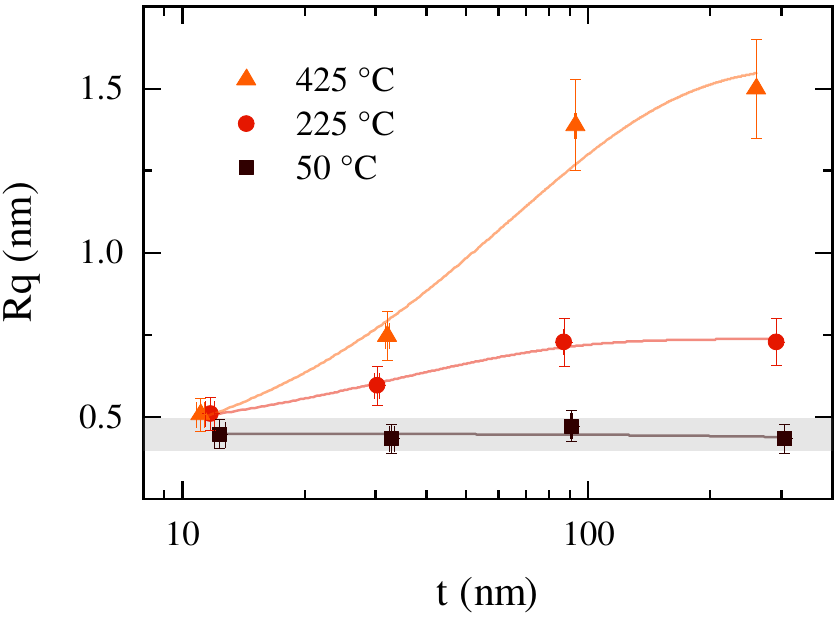}  
\caption{RMS roughness $Rq$ of $a$-Si films as a function of thickness t for samples grown at 50 \degree C (brown squares), 225 \degree C (red circles) and 425 \degree C (orange triangles). All samples are grown at 0.5 \si{\angstrom}/s. The gray area shows the substrate RMS roughness. Lines are guides to the eye.}
\label{Figure2b}
\end{figure}

The root mean square roughness $Rq$ is shown in Fig.~\ref{Figure2b} as a function of thickness for films grown at 50 \degree C, 225 \degree C and 425 \degree C. $Rq$ is less than 0.5 nm, the same as the roughness of the substrate, for films grown at 50 \degree C at all thicknesses. At higher growth temperature, $Rq$ increases proportionally to growth temperature and with thickness up to $\sim100$ nm, where it plateaus or slowly increases. A similar dependence of the surface roughness with thickness has been reported in hydrogenated $a$-Si films grown above room-temperature~\cite{Bruggemann2003ThicknessSilicon, Smets2003TemperatureGrowth}.

\subsection{Atomic density} 
\begin{figure}
\includegraphics[scale=0.95]{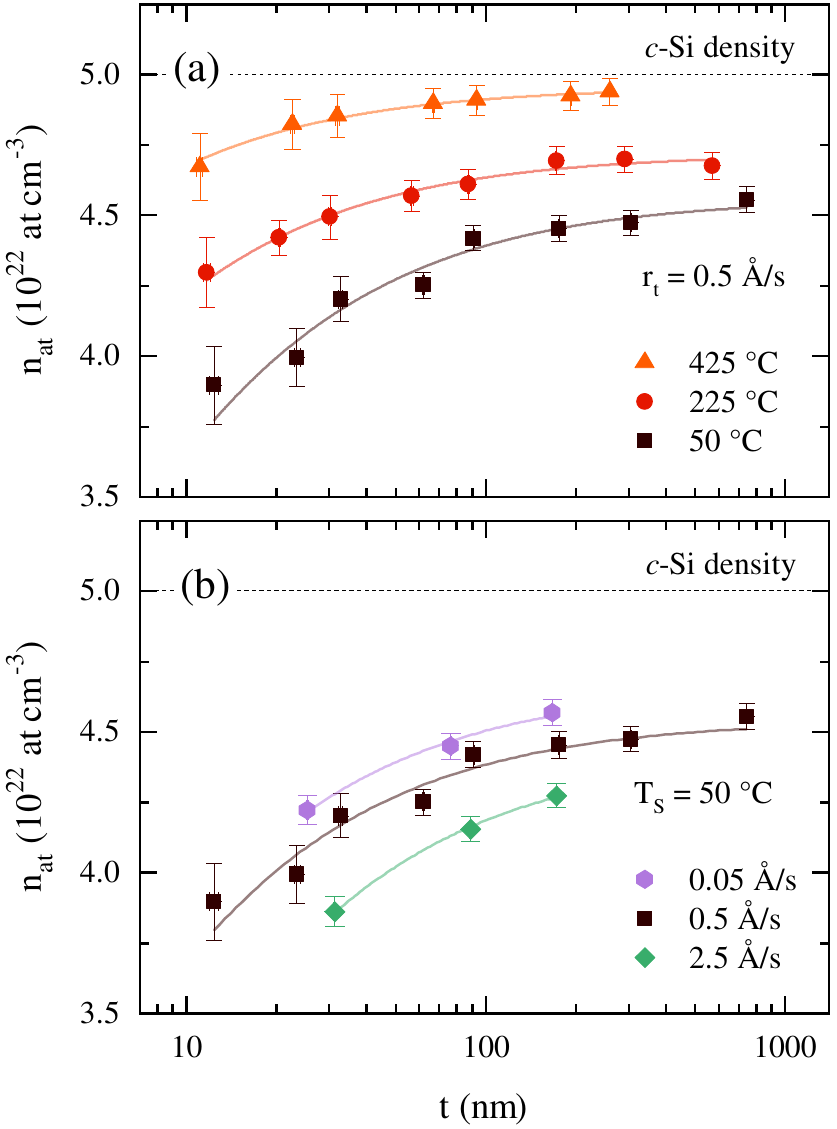}
\caption{Atomic density $n_{at}$ of $a$-Si films as a function of thickness t for (a) growth temperatures of 50 \degree C (brown squares), 225 \degree C (red circles) and 425 \degree C (orange triangles), and (b) growth rates of 0.05 \si{\angstrom}/s (lilac hexagons), 0.5 \si{\angstrom}/s (brown squares) and 2.5 \si{\angstrom}/s (green diamonds). In (a), all films are grown at 0.5 \si{\angstrom}/s, and in (b), all films are grown at 50 \degree C. In both plots the c-Si density is given for reference (dashed horizontal line). Solid lines are guides to the eye.}
\label{Figure1}  
\end{figure}

Atomic density $n_{at}$ of $a$-Si films grown at different temperatures and rates was determined by RBS measurements in combination with profilometry, where the thicknesses were corrected after the AFM measured roughness, as discussed in~\ref{density}. The resulting data are shown in Fig.~\ref{Figure1} as a function of thickness, which shows a remarkable and systematic dependence on thickness, growth temperature and rate. Error bars are dominated by the thickness uncertainty (profilometry) for thin films, and by areal atomic density uncertainty (RBS) for thick films. No observable discontinuities are seen. The lowest density films are quite underdense, $\sim22\%$ less than c-Si, whereas the highest density films are $\sim1\%$ less dense than c-Si. Previous work on vapor deposited $a$-Si films reported densities 19\% smaller than that of c-Si~\cite{Chittick1970PropertiesSilicon}, which was there suggested to be caused by by low-density regions or nanovoids found in $a$-Si~\cite{Moss1969EvidenceSilicon}.

Atomic density of $a$-Si films grown by e-beam is thus highly dependent on thickness, growth temperature and rate. We report density reductions with thickness for $a$-Si samples grown at room-temperature between 25\% and 10\% for the thinnest and thickest films, respectively, when compared to c-Si density. Higher growth temperature and lower growth rate yield films with higher atomic density at any given thickness. Thick $a$-Si films, around 600 nm thick, and grown at different temperatures were subsequently annealed at 425 \degree C in UHV for 3 hours. The film grown at 50 \degree C shows a thickness reduction from 598.6 nm to 593.1 nm ($-5.5\pm2.1$ nm, or -0.92\%), whereas the thickness variation for films grown at 225 and 425 \degree C is within error bars. Annealing at temperatures up to 425 \degree C thus does very little to the atomic density of $a$-Si films.

Atomic density provides a proxy to evaluate the thermodynamic and kinetic stability of amorphous solids~\cite{Ediger2017Perspective:Glasses}, where it is known that stability can be tuned by means of growth temperature and rate on vapor-deposited glasses~\cite{Berthier2017OriginGlasses}. Increasing growth temperature increases surface mobility during deposition and allows atoms to reach lower energy positions before being deposited over~\cite{Shi2011PropertiesFilms}. Similarly, slower growth rates allow these positions to be reached before mobile atoms are deposited over. In the particular case of $a$-Si and for the data presented in this work, atomic density shows a stronger dependence on growth temperature and thickness rather than growth rate. For this reason, we focus the present study on growth temperature and thickness dependence.

Resonant Rutherford backscattering spectrometry was performed on the films to determine the oxygen content through the samples. Films grown at 50 and 425 \degree C show 4\% and 1\% oxygen content, respectively, at all depths below the native oxide layer. Oxygen content did not increase with film aging (after 4 months). Neither contaminants nor water were observed via RBS and HFS, respectively, in the films. Thickness measurements were also performed both immediately and four months after deposition, no changes were observed in thickness as a function of time.

\subsection{Short- and Medium-range order}
\label{microscopyR}
In this section we present electron microscopy results, specifically diffraction, EELS, FEM, and high-resolution imaging results for $\sim30$ nm thick $a$-Si samples grown at 0.5 \si{\angstrom}/s, and at 50 \degree C, 250 \degree C and 450 \degree C.

\begin{figure}
\includegraphics[scale=0.95]{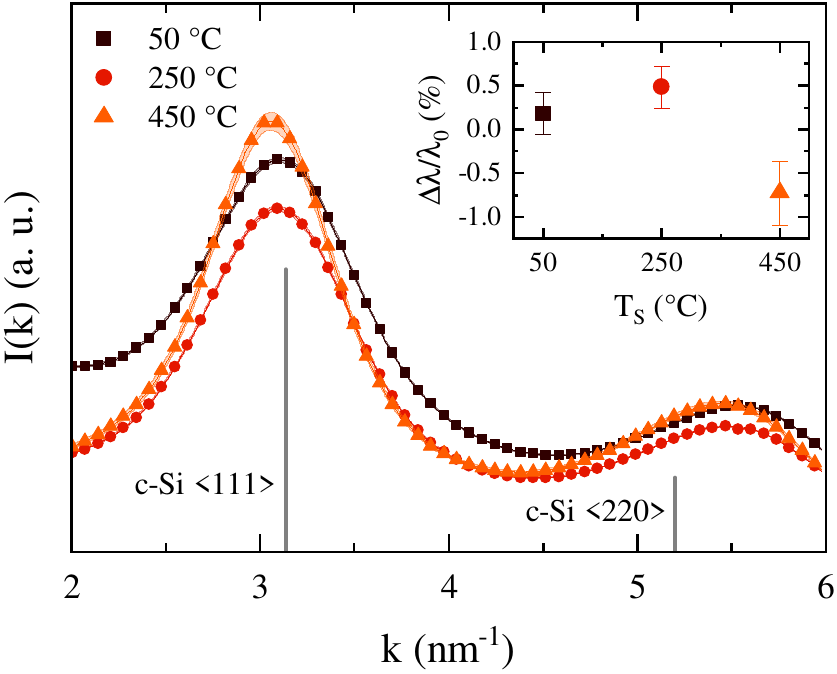}
\caption{Diffracted electron intensity $I(k)$ for $a$-Si thin films, $\sim30$ nm thick, grown at 0.5 \si{\angstrom}/s, and at 50 \degree C (brown), 250 \degree C (red) and 450 \degree C (orange). The diffraction peaks of c-Si are shown for reference. Inset: Bulk plasmon relative change $\Delta\lambda/\lambda_0$ from EELS as a function of growth temperature T$_\textrm{S}$ for the same samples. No significant change in average bond length between the films is observed.}
\label{Figure3}
\end{figure}

Figure~\ref{Figure3} shows the average diffracted intensities $I(k)$. ($k$ defined as $k=\theta /\lambda$, where $\lambda$ is the electron wavelength.) The peak positions are typical of $a$-Si and do not shift significantly from sample to sample. The first broad peak sits at $3.08 \pm 0.05$ nm$^{-1}$, which matches the c-Si $\langle111\rangle$ peak at 3.13 nm$^{-1}$. The second $a$-Si peak position at $5.54 \pm 0.05$ nm$^{-1}$ sits  between the c-Si reflections $\langle220\rangle$ at 5.2 nm$^{-1}$ and $\langle311\rangle$ at 6.1 nm$^{-1}$. None of the data show the sharp peaks that would indicate nanocrystallization. Isolated nanocrystals were observed in the 450 \degree C film, indicating that under these growth conditions crystallization of $a$-Si starts at a temperature between 425 \degree C and 450 \degree C. Electron microscopy characterization was performed between the nanocrystals on fully amorphous regions. Differences in peak heights between films are not well quantified in this data. For more quantitatively accurate $I(k)$ and structure factor $S(q)$, synchrotron experiments are needed. Similarities between positions and widths for both low and high-$k$ peaks seen in Fig.~\ref{Figure3}, however, suggest that there are no significant differences in SRO for the films grown at different temperatures, consistent with previous work on $a$-Si~\cite{Voyles2002FluctuationSTEM}.

EELS measurements performed on the same samples find an average bulk plasmon peak energy $\lambda_0$ of 16.6 eV. The relative change in $\lambda_0$, $\Delta\lambda/\lambda_0$ is +0.2\%, +0.5\%, and -0.7\% for samples grown at 50 \degree C, 250 \degree C and 450 \degree C, respectively (Fig.~\ref{Figure3} inset). At constant composition and temperature, $\Delta\lambda/\lambda_0 \propto \sqrt{\Delta \rho / \rho_0}$, so the maximum change in density, not counting contributions from voids, is negligible. Together, these results show that the average bond length for atoms within the network is the same for the different samples. The SRO structure, specifically bond length and coordination, is independent of the growth conditions, and there is no change in atomic number density except what is created by introducing voids, to be discussed below.

\begin{figure}
\includegraphics[scale=0.95]{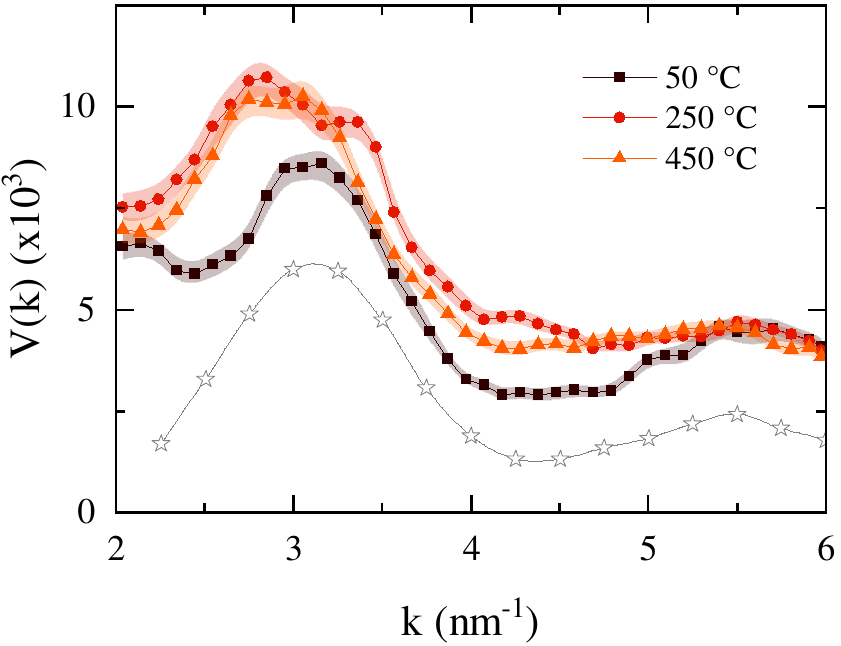}  
\caption{FEM data $V(k)$ as a function of wavevector $k$ of $a$-Si thin films, $\sim30$ nm thick, grown at 0.5 \si{\angstrom}/s, and at 50 \degree C (brown squares), 250 \degree C (red circles) and 450 \degree C (orange triangles). Grey stars show MRO typically observed in $a$-Si films from Ref.~\cite{Voyles2003Medium-rangeMicroscopy}. The data for the 250 \degree C and 450 \degree C films, however, reveal a MRO structure not previously reported.}
\label{Figure-FEM}  
\end{figure}

FEM data of the same films shown in Fig.~\ref{Figure3} are shown in Fig.~\ref{Figure-FEM}. $V(k)$ peaks in amorphous materials typically occur at the same $k$ as peaks in $I(k)$. That is the case for the film grown at 50 \degree C, but the films grown at 250 \degree C and 450 \degree C show a splitting of the first peak into a contribution near 3.09 nm$^{-1}$ and another peak at lower $k$, closer to 2.6 nm$^{-1}$. This peak indicates that local structures exist in the higher substrate temperature samples with a larger interatomic spacing than has been observed in any previous FEM experiments on $a$-Si. There are no larger interatomic spacings in diamond structure Si, but a different crystal structure for Si, Si24, which has diffraction features at lower $k$ corresponding to the peak position in $V(k)$~\cite{Kim2015SynthesisSilicon}. The low-$k$ diffraction in Si24 arises from 8-membered rings. In that work~\cite{Kim2015SynthesisSilicon}, the Si24 crystal was synthesized at high pressure with Na atoms embedded in the crystal, filling the 8-atom Si rings. The material was brought to room pressure, and the Na atoms were removed using thermal degassing. The resulting crystal is stable at room-temperature and atmospheric pressure. We do not suggest that there are nanocrystals of Si24 in the material, or even that we have created a paracrystalline analog to Si24. Instead we suggest that the low-$k$ feature in $V(k)$ of these $a$-Si films grown at higher temperatures (225 \degree C and 425 \degree C) could be caused by structures in the amorphous network involving large rings of 8 atoms or more, significantly larger than those believed to typically be present in $a$-Si~\cite{Beeman1977VibrationalSemiconductors}. These could even be viewed as extremely tiny nanovoids.

\begin{figure}
\includegraphics[scale=0.20]{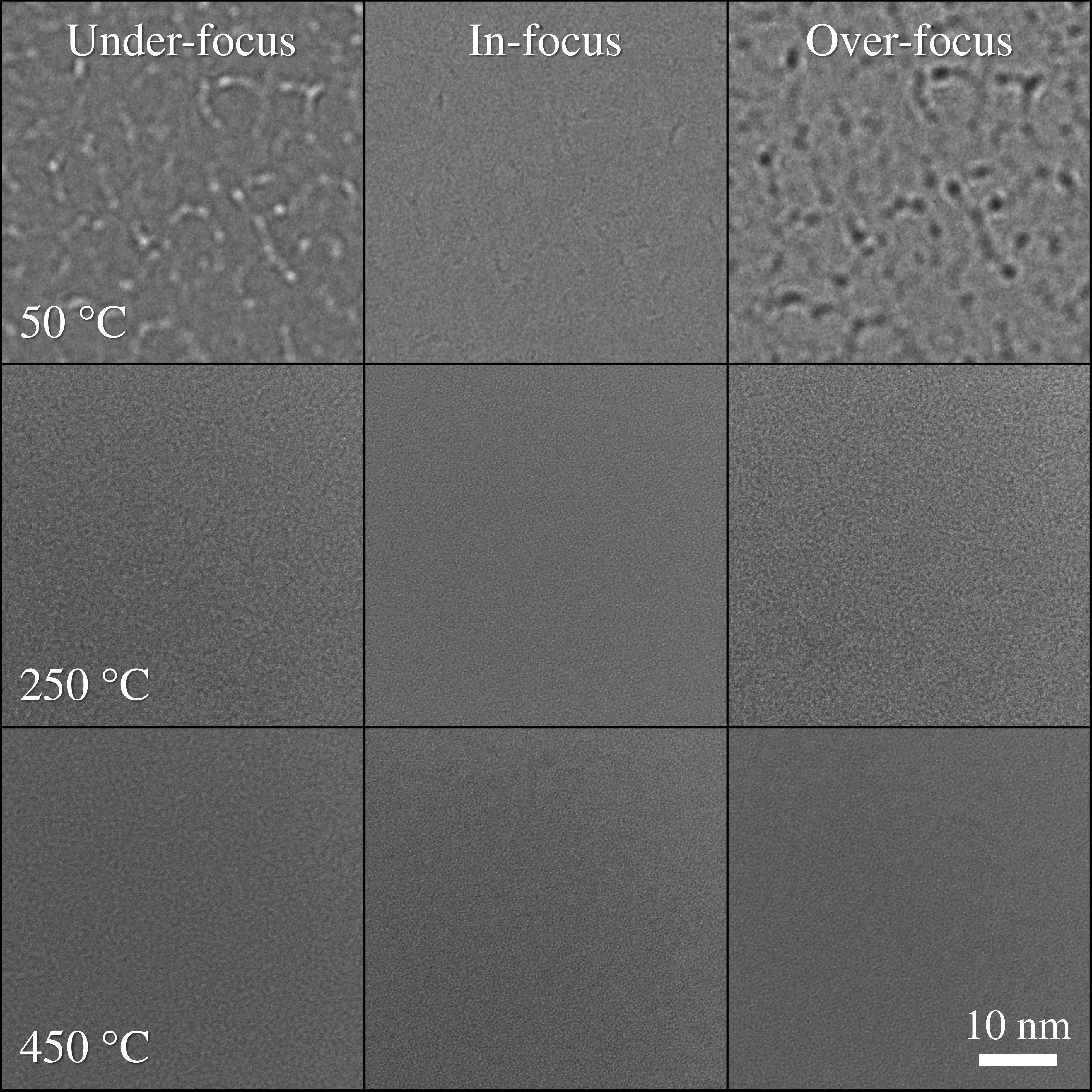}  
\caption{HRTEM images of $a$-Si thin films $\sim30$ nm thick grown at 0.5 \si{\angstrom}/s and at 50 \degree C, 250 \degree C and 450 \degree C (top to bottom row, respectively). Columns from left to right show under-focus, in-focus, and over-focus images (see description in text). The film grown at 50 \degree C shows columnar structure, visible as wiggly lines and voids visible as white features in the under-focus image and black features in the same place in the over-focus image. Films grown at higher temperatures show no notable features.}
\label{Figure4}
\end{figure}

Fig.~\ref{Figure4} shows under-focus, in-focus, and over-focus HRTEM images of the same samples. Under- and over-focus image pairs are a classical method for identifying voids in materials from the switch in contrast of the Fresnel fringe surrounding the void. The 50 \degree C sample shows clear evidence of columnar microstructure with nanometer-scale voids forming part of the column boundaries. The higher temperature films do not have the columnar microstructure and do not have voids detectable by this method. In our previous work~\cite{Queen2015JNCS}, cross-sectional TEM images (XTEM) of $a$-Si films shown a columnar structure whose diameter increases with growth temperature. Such structure is common in amorphous films grown by PVD techniques~\cite{Bales1991MacroscopicDeposition}, including e-beam evaporation~\cite{Hulsen1996SurfaceFilms}. These images also reveal a region close to the substrate interface that lacks columnar microstructure; this region is thicker for higher growth temperature. The columnar structure observed for the film grown at 50 \degree C has not developed yet for films grown at 250 \degree C and 450 \degree C; therefore, the thickness at which columns develop in $a$-Si thin films is growth temperature dependent.

\subsection{Bond angle deviation and local strain}
\begin{figure}
\includegraphics[scale=0.95]{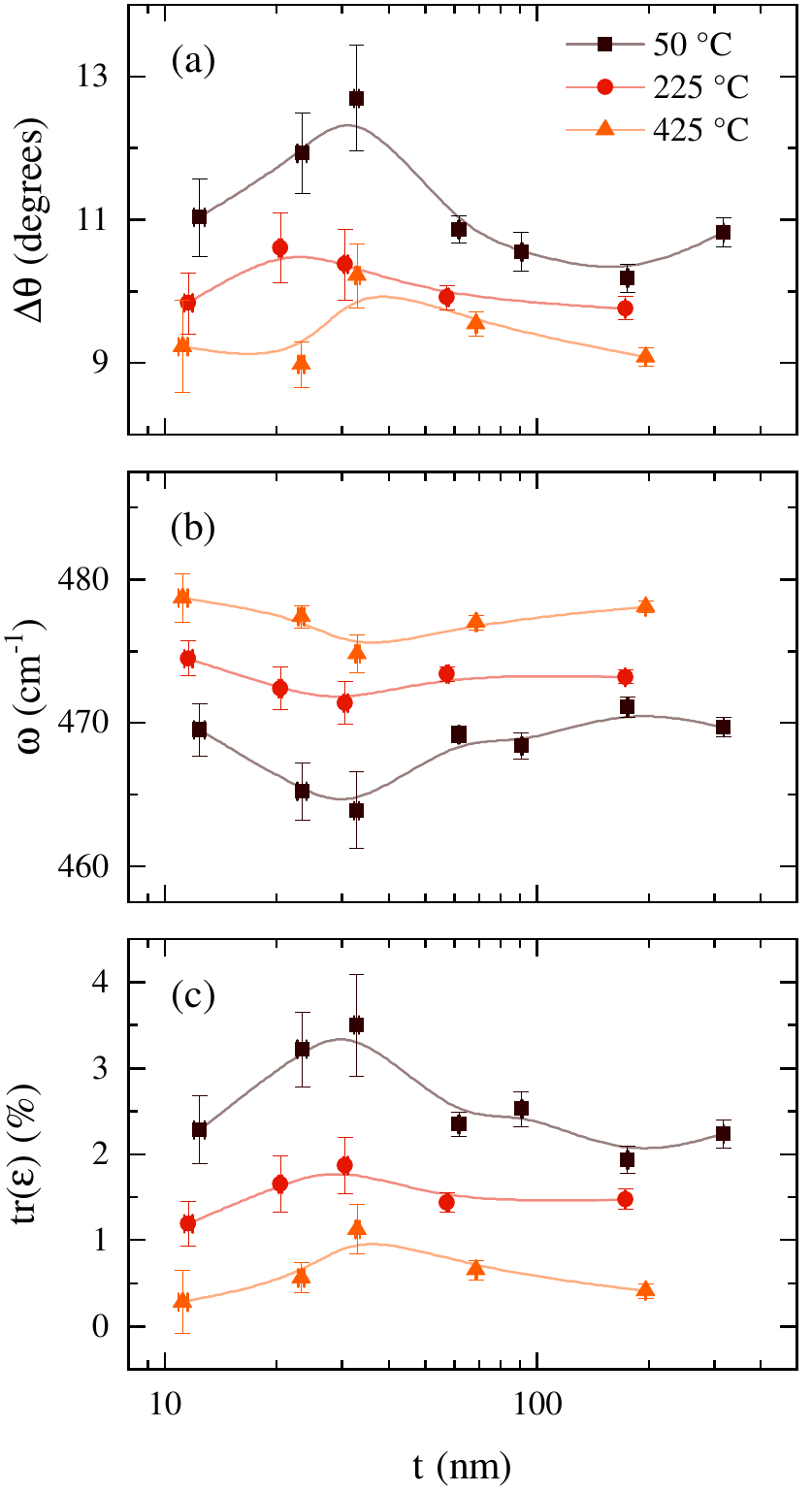}  
\caption{(a) Bond angle deviation $\Delta\theta$, (b) TO peak position $\omega$, and (c) trace of strain $tr(\epsilon)$ as a function of thickness t for samples grown at 50 \degree C (brown squares), 225 \degree C (red circles) and 425 \degree C (orange triangles). Lines are guides to the eye.}
\label{Figure5}  
\end{figure}

Electron microscopy was used as described above to characterize SRO. $I(k)$ and $\Delta\lambda/\lambda_0$ are mostly sensitive bond length and coordination number, and less to bond angles. We here turn to Raman to characterize the deviations from the tetrahedral angle $\theta$ that are found in $a$-Si. Figure~\ref{Figure5}(a) shows the bond angle deviation $\Delta\theta$ as a function of thickness for a series of growth temperatures. The range of $\Delta\theta$ found for these samples, approximately 9\degree~to 13\degree, is similar to values reported for $a$-Si model structures~\cite{Beeman1985StructuralSilicon}. Films grown at 425 \degree C show $\Delta\theta$ that ranges from 9\degree~to 10\degree. These values fall on the low end of those obtained from the radial distribution function of $a$-Si~\cite{Fortner1989RadialSilicon}; and to our knowledge are amongst the lowest reported by experiment. Note that $\Delta\theta = 0\degree$ for c-Si, and has been shown theoretically to be $\geq 6.6\degree$ for $a$-Si~\cite{Beeman1985StructuralSilicon}, evidence that the transition from crystalline to amorphous structures is not continuous. $\Delta\theta$ is considered a measure of disorder at very short length scales; nearest neighbor distances, or around 2 \AA. At each thickness, $\Delta\theta$ is lower for films grown at higher temperatures, indicating that disorder decreases with increasing growth temperature.

Fig.~\ref{Figure5}(b) shows the TO peak position $\omega$ and Fig.~\ref{Figure5}(c) shows the trace of the local strain $tr(\epsilon)$ calculated from $\omega$ as described in~\ref{BADM}. $\Delta\theta$ and $tr(\epsilon)$ shown in Figs.~\ref{Figure5}(a) and \ref{Figure5}(c), respectively, are correlated. Higher growth temperature thus yields less disordered and less strained films at all thicknesses. However, the monotonic trend observed for atomic density as a function of thickness is not seen for these properties: both $\Delta\theta$ and $tr(\epsilon)$ increase with thickness up to a maximum value, then decrease. This behavior is most clear in films grown at 50 \degree C, where the maximum values of $\Delta\theta$ and $tr(\epsilon)$ occur near 30 nm. For films grown at higher temperature, the peak in $\Delta\theta$ and $tr(\epsilon)$ is more subtle, and may occur at a different thicknesses.

Raman data analysis shows a relaxation process in which disorder and local strain reduce above a \textit{critical thickness}; however, it is unclear whether this process implies a structural change in the atoms underneath, i.e., the \textit{reorganization} of the atoms already deposited.

\begin{figure}
\includegraphics[scale=0.95]{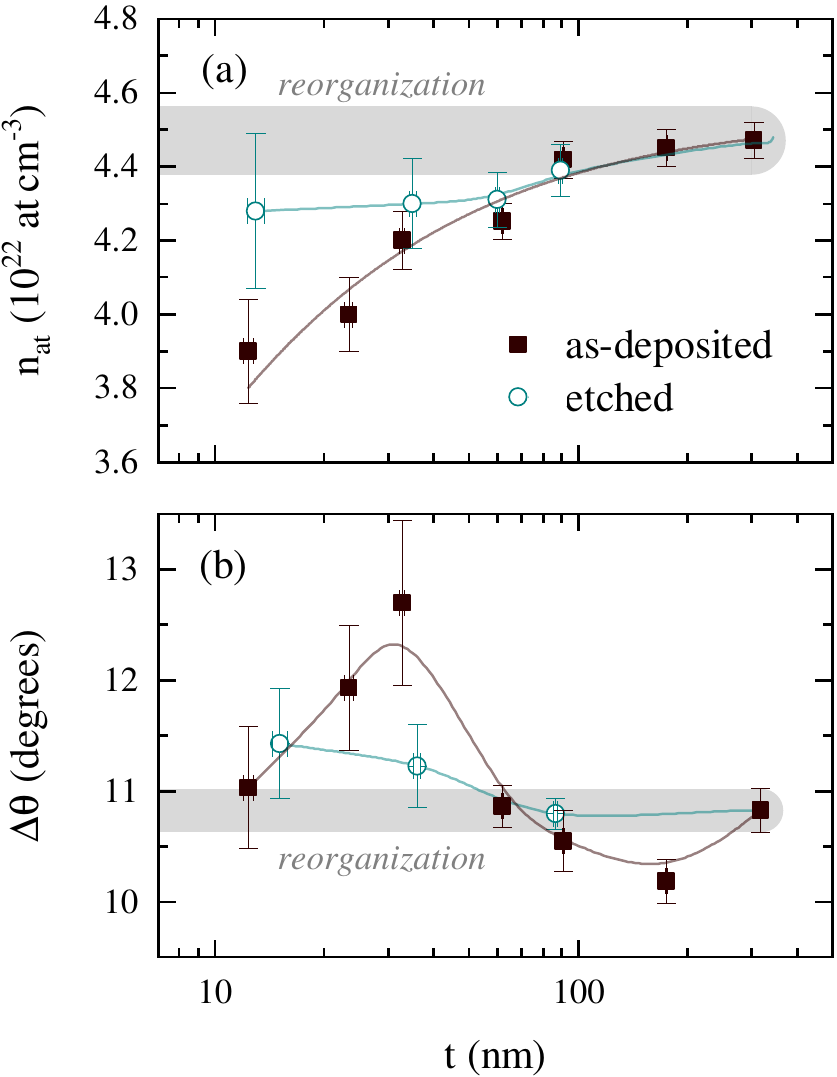}  
\caption{(a) Atomic density $n_{at}$, and (b) bond angle deviation $\Delta\theta$ as a function of thickness t for as-deposited (brown closed squares) and etched (cyan open circles) films. All films were grown at 50 \degree C and 0.5 \si{\angstrom}/s; etched films initially were $\sim$\,300 nm thick. Gray areas labeled `reorganization' show the extrapolated $n_{at}$ and $\Delta\theta$ values of the $\sim300$ nm as-deposited films prior to the etch process (see text for further details). Lines are guides to the eye.}
\label{Figure6}
\end{figure}

Following the procedure described in~\ref{methods}, we study $n_{at}$ and $\Delta\theta$ of several $a$-Si etched films. In Fig.~\ref{Figure6} we compare these results to equivalent as-deposited films to probe for reorganization during growth. The larger error bars are due to increased surface roughness and thickness uncertainty after the etch process. All films etched and measured were grown at 50 \degree C because they exhibit the largest change in atomic density and bond angle deviation as a function of thickness, as seen in Figs.~\ref{Figure1}(a) and \ref{Figure5}(a), respectively. The gray areas labeled `reorganization' in Fig.~\ref{Figure6} are the extrapolated $n_{at}$ and $\Delta\theta$ values of thick as-deposited films. A complete reorganization would yield etched thin films with the higher $n_{at}$ and lower $\Delta\theta$ of the as-deposited thicker films, whereas a lack of reorganization would yield etched thin films with the lower density and higher disorder of the as-deposited thin films.

Fig.~\ref{Figure6}(a) shows that atomic density $n_{at}$ from etched films is higher than that of as-deposited films with similar thickness, most consistent with reorganization, with etched films lying within error bars of the as deposited values of thick films and well above the values of as-deposited thin films. Bond angle deviation $\Delta\theta$ from etched films is lower than that of as-deposited films with similar thickness, shown in Fig.~\ref{Figure6}(b), which suggests a reduction of disorder, and are also most consistent with reorganization of the early layers of atoms deposited near the substrate.

It has been proposed that due to the large interface energy between substrate and film, thin films prepared by physical vapor deposition grow via a Volmer-Weber mode~\cite{Floro2001TheFilms, Floro2002PhysicalFilms}. This proposed growth mode is in agreement with stress measurements in e-beam evaporated $a$-Si films, which show the nucleation of columns that correlate with the transition from compressive to tensile stress regimes as a function of thickness~\cite{Floro2003OriginsFilms}.

The results presented in this section show structural relaxation as a function of thickness that leads to the reduction of disorder and local strain, compatible with the previously described studies of the stress evolution during growth of evaporated $a$-Si~\cite{Floro2003OriginsFilms}. From this work~\cite{Floro2003OriginsFilms}, we also see that the dependence with thickness of the compressive-to-tensile stress transition is in agreement with our structural observations by XTEM, which show the onset of the columnar growth~\cite{Queen2015JNCS}. The stress of $a$-Si films grown at room-temperature was measured through the curvature of their substrates reporting tensile stresses of $694\pm94$ MPa for a 30 nm film, and $523\pm20$ MPa for a 90 nm film.

\subsection{Transverse sound velocity}
\begin{figure}
\includegraphics[scale=0.95]{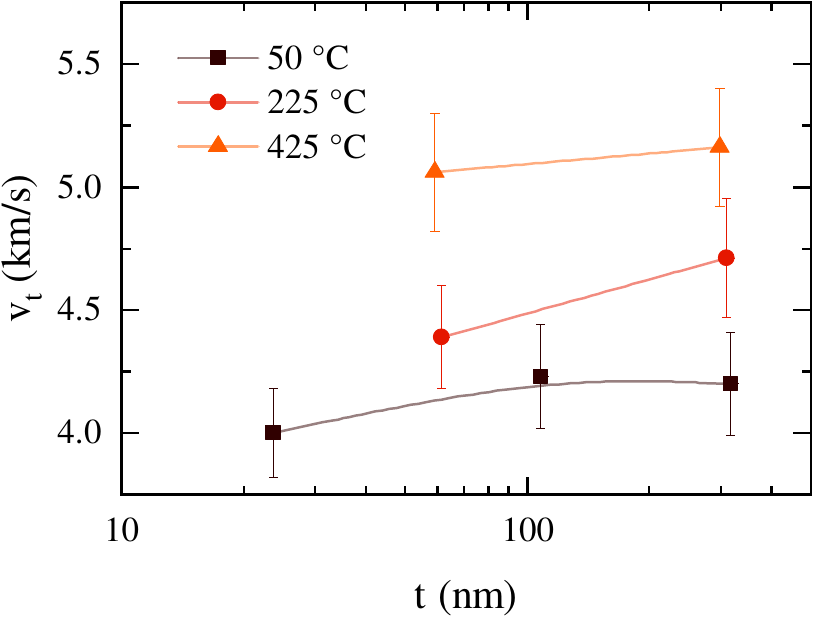}  
\caption{Transverse sound velocity $v_{t}$ as a function of thickness t of $a$-Si films grown at 50 \degree C (brown squares), 225 \degree C (red circles) and 425 \degree C (orange triangles). Lines are guides to the eye.}
\label{Figure11}  
\end{figure}

Transverse sound velocity $v_{t}$ as a function of thickness for films grown at different temperatures is shown in Fig.~\ref{Figure11}. Sound velocity is only weakly dependent on thickness, perhaps within error bars of constant, whereas it increases with increasing growth temperature.

Sound velocity in solids, including amorphous solids, is due to phonons that propagate through the material, with a velocity that is independent of the frequency of the phonons for relatively long wavelength phonons. The sound velocity is generally dependent on the interatomic spacing, the atomic mass, and the interatomic bond strength. In $a$-Si, the increase of sound velocity $v_t$ with increasing growth temperature, shown in Fig.~\ref{Figure11}, correlates with the reduction of atomic disorder ($\Delta\theta$) and local strain ($tr(\epsilon)$), shown in Figs.~\ref{Figure5}(a) and \ref{Figure5}(c), respectively. That sound velocity depends strongly on growth temperature, but is not strongly dependent on thickness at constant growth temperature, despite significant changes in atomic density with thickness, shows that sound waves are carried through an $a$-Si network. Sound velocity, therefore, is not affected directly by the overall density, i.e., it is not much affected by nanovoids.

\subsection{Dangling bond density}
\begin{figure}
\includegraphics[scale=0.95]{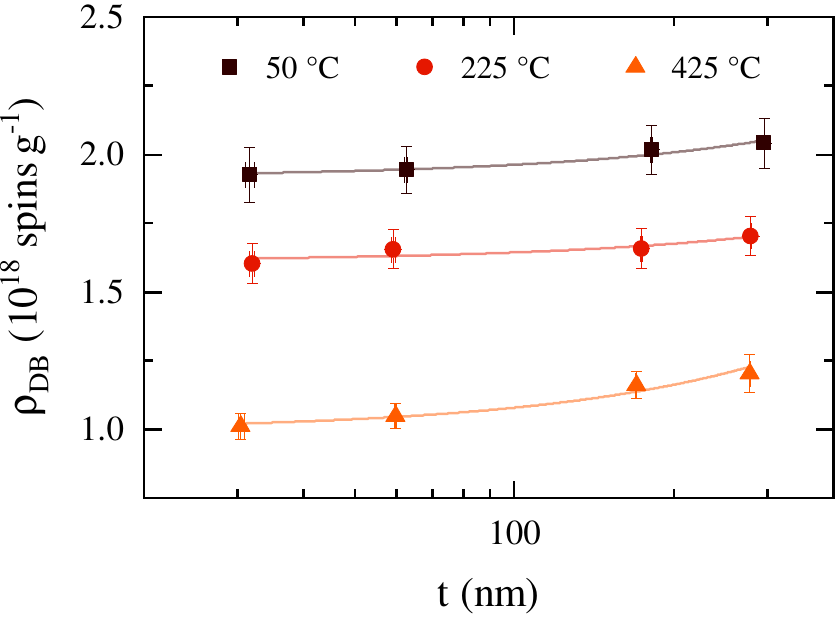}  
\caption{Dangling bond density $\rho_{DB}$ as a function of thickness t of $a$-Si films grown at 50 \degree C (brown squares), 225 \degree C (red circles) and 425 \degree C (orange triangles). All samples were grown at 0.5 \si{\angstrom}/s. Lines are linear fits to the data.}
\label{Figure10}  
\end{figure}

Dangling bond density $\rho_{DB}$ has long been considered a defect metric and is correlated with surface-state transitions in silicon~\cite{Rowe1973Surface-stateSpectra}. In photovoltaic and semiconductor technologies, $a$-Si is prepared with hydrogen in order to passivate dangling bonds and create high quality, or device quality, films. The dangling bond densities of our films are on the order of 10$^{18}$ spins/cm$^3$ (see Fig.~\ref{Figure10}). Standard $a$-Si has dangling bond densities $\sim10^{19}$ spins/cm$^3$, and device-quality $a$-Si:H shows values $<10^{16}$ spins/cm$^3$~\cite{Mahan1991}. We report $\rho_{DB}$ per unit mass (g$^{-1}$) rather than per unit volume (cm$^{-3}$) because the atomic density of our samples is not constant, which makes the former a more accurate metric for direct comparisons.

Figure~\ref{Figure10} shows that $\rho_{DB}$ decreases significantly with increasing growth temperature, and mildly increases with thickness for all growth temperatures. The former statement is consistent with the strong dependence of atomic density $n_{at}$ on growth temperature, but is the opposite to how $n_{at}$ depends on thickness, suggesting that the relationship between $n_{at}$ and $\rho_{DB}$ is not simple. The dependence of $\rho_{DB}$ on growth temperature suggests that the nucleation of dangling bonds is inversely proportional to surface diffusion; specifically, that higher growth temperature yields more 4-fold coordinated atoms. That dangling bond density mildly \textit{increases} with thickness for all growth temperatures, suggests that the formation of this type of defect is not related to the mechanisms responsible for the films' atomic density or the atomic reorganization process previously discussed.

The short- and medium-range order, bond angle deviation, local strain and dangling bond density results report information about the distribution of silicon atoms in the films and about specific electronic defects; specifically, EELS results show that the interatomic distances do not change with growth temperature. Therefore, these results do not explain the very low atomic density values, and particularly that of the thinnest films grown at room-temperature. In the next section, we study the presence of nanovoids, which in $a$-Si cannot be detected by microscopy techniques, but are considered a common structural defect in $a$-Si~\cite{Guerrero2020ComputationalDensity}.

\subsection{Nanovoids characterization}
\label{nanovoidsR}
\begin{figure}
\includegraphics[scale=0.95]{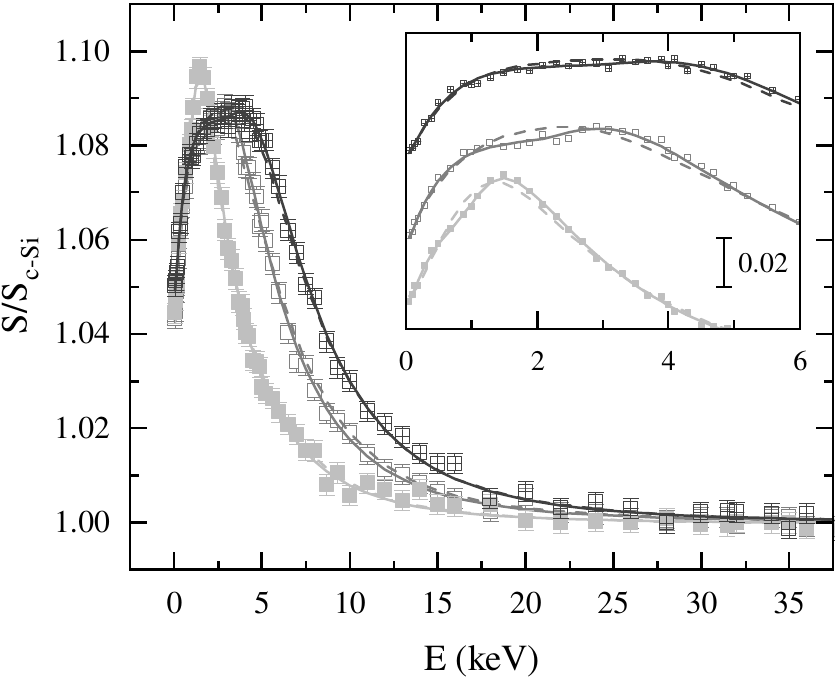}  
\caption{S-parameter normalized to the c-Si S-parameter (S$_\textrm{{c-Si}}=0.5324$) as a function of incident positron energy E of three $a$-Si films grown at 50 \degree C and 0.5 \si{\angstrom}/s, with thicknesses of 66 nm, 183 nm and 308 nm (light to dark grey, respectively). Data at 0 keV are values from the $a$-Si surface, and move progressively through the film for increasing energies until the beam reaches the substrate at $\sim30$ keV. Inset: detail at low energies showing the accuracy of the single (dashed lines) and dual (solid lines) models. The dual model improves its relative accuracy with respect to the experimental data almost 50\% compared to the single model. Curves have been vertically shifted to better show the fittings.}
\label{Figure7}  
\end{figure}

Doppler broadening spectroscopy results for films of different thicknesses and grown at 50 \degree C are shown in Fig.~\ref{Figure7}, where the S-parameter is plotted as a function of energy. Energy is proportional to penetration depth (as described in~\ref{nanovoidsM}) and, for the films reported in this work, it ranges from 0 keV (surface) to $\sim30$ keV (substrate). We could not measure films thinner than 60 nm due to lack of sensitivity. The data acquired from the annihilation of positrons confirm that all $a$-Si films measured, grown at different temperatures and for different thicknesses, contain nanovoids. However, the S-parameter, which is proportional to the empty space volume, is not calibrated for $a$-Si limiting our ability to quantify the size of its nanovoids. Electron microscopy cannot detect the presence of nanovoids in $a$-Si films due to the atomic superposition and the low atomic scattering factor of silicon. For these reasons, we can only report a qualitative analysis of the data and establish relative trends for the empty space, or nanovoids, volume measured in these $a$-Si films.

We used two different models to simulate the experimental data: 1) a \textit{single} volume of empty space that is uniformly distributed throughout the film (one S-parameter), and 2) a \textit{dual} volume of empty space that are uniformly distributed throughout the film (two S-parameters), in which the nanovoids volume closer to the substrate is different than that closer to the surface. The simulations yield better results for the dual model, which is almost 50\% more accurate than the single model (see inset in Fig.~\ref{Figure7}). The dual volume of empty space is the simplest model to probe whether the data suggests a more complex distribution of nanovoids than a single volume of empty space model. These results suggest that, at least, two different distributions of nanovoids are present on $a$-Si films thicker than 60 nm, with the interface between the two distributions at $31\pm6$ nm from the substrate for films grown at 50 \degree C, and at $59\pm17$ nm for films grown at 425 \degree C. This increase of the first nanovoids layer thickness with growth temperature is in agreement with the thickness dependence of void tracks formation with growth temperature seen in $a$-Si films grown by Floro et al.~\cite{Floro2003OriginsFilms}. The dual model also reports that the first layer, closer to the substrate, has a larger volume of empty space than the second layer, closer to the surface. However, since the S-parameter to nanovoid size calibration is not available, we cannot quantify the volume reduction of empty space between the two distributions and the relative change in size and concentration of nanovoids.

\begin{figure}
\includegraphics[scale=0.95]{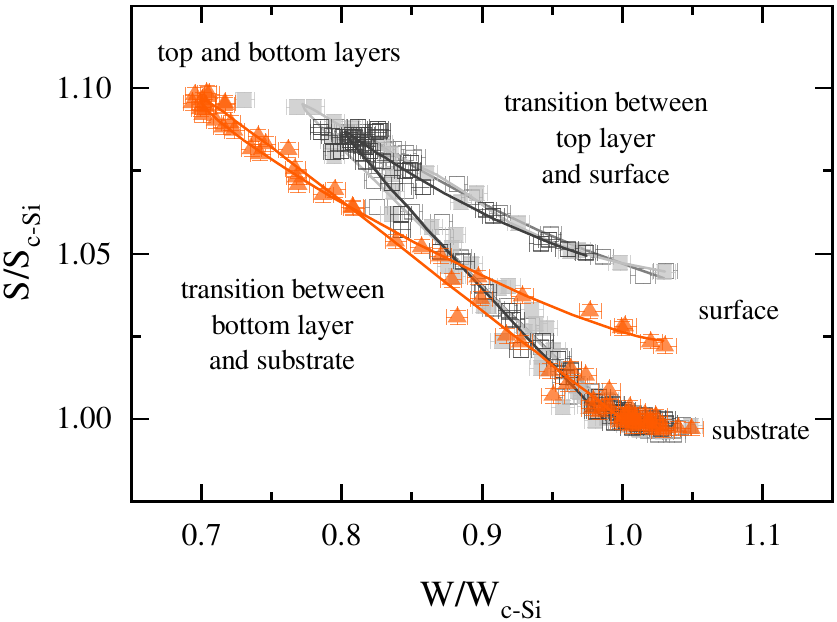}  
\caption{$S$ vs $W$ plots for films grown at 50 \degree C, with 66 nm, 183 nm and 308 nm (light to dark grey squares, respectively), and for a film grown at 425 \degree C with 291 nm (orange triangles). Vertical and horizontal axes are normalized to the c-Si $S$ and $W$ values, with S$_\textrm{{c-Si}}=0.5324$ and W$_\textrm{{c-Si}}=0.0237$. Solid lines are $S$-$W$ fits considering a bilayer distribution of nanovoids. Labels indicate the position where positrons annihilate from film surface to substrate.}
\label{Figure8}  
\end{figure}

$S$ vs $W$ plots are shown in Fig.~\ref{Figure8} for the same films shown in Fig.~\ref{Figure7} and a $\sim300$ nm film grown at 425 \degree C. Typical $S$ vs $W$ plots show annihilated positrons with a constant $S/W$ ratio from surface to substrate (a single straight line). Our results, however, show a ``V'' shaped line due to the presence of different types of nanovoids across the film (nanovoids of different size). In other words, different $S$-$W$ pairs, or slopes, represent nanovoids with a chemically distinct surface bonding. Steeper $S$-$W$ slopes correspond to larger nanovoids. Points at the end of the ``V'' with moderate $S$ and high $W$ values exist at the surface of the film, and progressively change by following the ``V'' shape to the other end with low $S$ and high $W$, where positrons annihilate at the interface between the two distributions of nanovoids (see Fig.~\ref{Figure8} and corresponding labels). The chemical change of the nanovoids across the distinct distributions is due to the distinct chemical nature of electrons present in them, i.e., to the nanovoids sizes and not to their concentration.

The analysis of the $S$ data suggests that a single distribution is less likely than a dual distribution of empty space volume, or total nanovoids volume, as shown in Fig.~\ref{Figure7} inset, and the $S$-$W$ data in Fig.~\ref{Figure8} shows more than one type of nanovoid, which significantly differ in the nature of the bonding at their inner surfaces. Films grown at all temperatures show that the distribution closer to the surface has a smaller empty space volume than the distribution closer to the surface. Additionally, the empty space volume in the distribution closer to the surface is smaller for films grown at 425 \degree C compared to films grown at 50 \degree C.

In $a$-Si, dangling bonds are found to occur on the inner surfaces of nanovoids \cite{Connell1976UseGermanium, Knights1979DefectsA-Si:H, Street1980LuminescenceSilicon, Dabrowski1994AtomicExperiment}. However, the relationship in the literature between dangling bonds and nanovoids is unclear. 

The existence of nanovoids in $a$-Si, confirmed by DBS measurements, provides an explanation for the the low atomic density values of the films reported in this work.

\section{Discussion and conclusions}
Atomic density $n_{at}$ of $a$-Si films increases with increasing growth temperature and thickness, and with reducing growth rate (Fig.~\ref{Figure1}). In all cases films are underdense, with $n_{at}$ as high as $\sim99\%$ of c-Si density for films grown at 425 \degree C and thicker than 300 nm, and as low as $\sim78\%$ of c-Si density, which is achieved for the thinner films grown at room-temperature. FEM results on $\sim30$ nm thick films show significant differences in MRO, as shown in Fig.~\ref{Figure-FEM}, despite no change in SRO (Fig.~\ref{Figure3}), which suggests that larger membered rings appear as the growth temperature increases. Measurements of the plasmon peak energy relative change $\Delta\lambda/\lambda_0$ presented in the inset of Fig.~\ref{Figure3}, show no significant changes in atomic density between films as a function of growth temperature. Therefore, the reduction of $n_{at}$ can only be attributed to voids. Those voids must be nanometer-scale, as large voids would be visible in HRTEM.

Raman data analysis reveals a reduction of the atomic disorder, the bond angle deviation $\Delta\theta$, with increasing growth temperature [Fig.~\ref{Figure5}(a)]. This dependence on growth temperature is also observed for the position of the transverse-optic peak [Fig.~\ref{Figure5}(b)], which correlates with the local strain $tr(\epsilon)$ of the $a$-Si films [Fig.~\ref{Figure5}(c)]. These data show a non-monotonic behavior of disorder and local strain with thickness, particularly for films grown at room-temperature, whose magnitudes build up and then reduce to their initial values. The characterization of nanovoids shows that the best simulations of the data (Fig.~\ref{Figure7}) and the analysis of the $S$-$W$ plots (Fig.~\ref{Figure8}) suggest the presence of at least two different types, or sizes, of nanovoids, with larger empty space volume near the substrate. The thickness at which the transition between the two distributions occurs, increasing from approximately 30 nm to 60 nm with increasing growth temperature, coincides with the critical thickness seen by Raman. The stress results also report higher tensile stress for thinner films, $\sim30$ nm thick, than for thicker films, $\sim90$ nm thick. We therefore speculate that at the critical thickness a structural relaxation process occurs, triggering the formation of columns observed by electron microscopy~\cite{Floro2003OriginsFilms, Queen2015JNCS}, and the reduction of $\Delta\theta$ (atomic disorder) and $tr(\epsilon)$ (local strain). Floro et al.~\cite{Floro2003OriginsFilms}, see similar effects, albeit at slightly different thicknesses.

The study of $n_{at}$ and $\Delta\theta$ from etched samples, Figs.~\ref{Figure6}(a) and \ref{Figure6}(b), respectively, shows reorganization of the films' structure during growth. These results do not suggest a direct relationship between reorganization and critical thickness, even though they might be caused by a common underlying relaxation process that takes place during growth. We note that the atomic density increases monotonically with thickness, suggesting that the reorganization of the films continuously happens during growth, whereas the critical thickness seems to be triggered by an specific event. Additionally, DBS results based on a dual distribution of empty space volume show that the total volume of nanovoids reduces as films grow thicker. Our data do not allow us to establish whether reorganization and critical thickness are two processes independent of each other, or whether one is triggered by the other.

Dangling bond density $\rho_{DB}$ results report up to $\sim1$ defect in $10^{4}$ atoms, which implies that Si atoms in $a$-Si films are largely fully coordinated despite their notably low atomic densities in some films, with up to $\sim22\%$ missing atoms (compared to c-Si). These results suggest that only a very small fraction (less than 1\%) of nanovoids contain even one dangling bond. Additionally, energetic considerations suggest that having two dangling bonds in a single nanovoid is unlikely because two nearby under-coordinated silicon atoms will bond together. Therefore, the increase of $\rho_{DB}$ with thickness (Fig.~\ref{Figure10}) suggests an increase of the nanovoid number, which in combination with the densification of the films (Fig.~\ref{Figure1}) imply that nanovoids become smaller as films grow thicker. This conclusion is also supported by DBS results (Fig.~\ref{Figure8}).

Two distinct processes happen in e-beam evaporated $a$-Si films during growth: i) a sudden structural relaxation at a \textit{critical thickness}, which depends on growth temperature and correlates with the formation of columns~\cite{Floro2003OriginsFilms, Queen2015JNCS} (Figure~\ref{Figure4}). This process is also captured by the reduction of atomic disorder and local strain (Figure~\ref{Figure5}), and by the reduction of nanovoids size and their total volume (Figure~\ref{Figure8}). And ii) a continuous \textit{reorganization} of the atoms seen in the atomic density and bond angle deviation of etched thicker films (Figure~\ref{Figure6}). This process is responsible for the reduction of the total volume of nanovoids as films grow thicker, which leads to the continuous densification of the films (Figure~\ref{Figure1}) and to the reduction of TLSs~\cite{Queen2013, Molina-Ruiz2021}.

As previously reported, the TLS density measured for $a$-Si depends strongly on growth conditions, with a dependence that can be completely explained by plotting TLS density as a function of atomic density. Remarkably, and as previously observed~\cite{Queen2013}, the excess heat capacity at low temperature, below 10 K, of thinner $a$-Si films (112 nm) is larger than that of thicker films (278 nm). This observation cannot solely be explained by the larger concentration of nanovoids in the first layer (near the substrate), which indeed would yield thinner films with larger excess specific heat, but not heat capacity. This excess heat capacity reduction can only be explained by atomic reorganization, which effectively reduces the number of TLSs when films grow thicker. Similarly, the mechanical loss of thinner $a$-Si films (59 nm) is larger than that of thicker films (299 nm)~\cite{Molina-Ruiz2021}. Our data suggest that the structural origin of TLSs in $a$-Si likely occur in highly disordered regions where nanovoids are present. The reduction of atomic disorder and nanovoids volume (increase in atomic density) correlates with the reduction of TLSs observed in thicker $a$-Si films. The reorganization of the atoms as films grow thicker plays a crucial role in reducing TLSs in $a$-Si, and it is likely to be related to local structural rearrangements caused by structural relaxation processes. The reduction of TLSs with growth temperature is likely to be related to surface diffusion mechanisms during growth, since surface diffusion is enhanced by temperature. Additionally, Fig.~\ref{Figure11} shows that sound velocity increases with increasing growth temperature, while its dependence on thickness is within error bars of no dependence, suggesting that the sound velocity depends on properties of the $a$-Si network, such as bond angle disorder, and not on the presence of nanovoids; while nanovoids and their environment, and not the constrained network, are responsible for the mechanisms that enable TLSs in $a$-Si.

\section*{Acknowledgements}
We thank P. Ci and J. Wu for assistance with Raman measurements; R. Chatterjee and J. Cooper for assistance with EPR measurements; D. Castells-Graells for assistance with films growths; and D. Strubbe for fruitful discussions on $a$-Si:H Raman spectra calculations. Samples preparation, ion beam analysis and Raman characterization were done at UCB supported by NSF (DMR-1508828 and DMR-1809498). Electron microscopy characterization was done at UW-Madison supported by the Wisconsin MRSEC (DMR-1720415). Sound velocity characterization was done at NRL supported by the Office of Naval Research. Electron paramagnetic resonance measurements were done at LBNL Joint Center for Artificial Photosynthesis supported by DOE (DE-SC0004993). Doppler broadening spectroscopy characterization was done at WSU supported by the late Dr. Kelvin Lynn and the Center of Materials Research (now Institute of Materials Research).

\bibliography{references}

\end{document}